\documentclass[10pt,twocolumn,letterpaper]{article}

\usepackage{iccv}
\usepackage{times}
\usepackage{amsthm,amsmath}
\usepackage[utf8]{inputenc} 
\usepackage{amsthm,amssymb,graphicx,cite}
\usepackage{color}
\usepackage{enumitem}
\usepackage{multirow}
\usepackage{siunitx}
\usepackage{booktabs}
\usepackage{balance}
\usepackage{times}
\usepackage{epsfig}
\usepackage{mathtools}
\usepackage{subcaption}
\usepackage{gensymb}
\usepackage{arydshln}

\usepackage[breaklinks=true,bookmarks=false]{hyperref}

\DeclareMathOperator{\dis}{d}

\iccvfinalcopy 

\def\iccvPaperID{****} 

\ificcvfinal\fi

\begin{document}


\title{Points2Sound: From mono to binaural audio using 3D point cloud scenes}

\author{Francesc Llu\'is\\
\and
Vasileios Chatziioannou\\
\and
Alex Hofmann
\and
Department of Music Acoustics (IWK)\\
University of Music and Performing Arts Vienna, Austria\\
{\tt\small lluis-salvado@mdw.ac.at}
}

\maketitle
\ificcvfinal\thispagestyle{empty}\fi

\begin{abstract}
For immersive applications, the generation of binaural sound that matches its visual counterpart is crucial to bring meaningful experiences to people in a virtual environment. Recent studies have shown the possibility of using neural networks for synthesizing binaural audio from mono audio by using 2D visual information as guidance. Extending this approach by guiding the audio with 3D visual information and operating in the waveform domain may allow for a more accurate auralization of a virtual audio scene. We propose Points2Sound, a multi-modal deep learning model which generates a binaural version from mono audio using 3D point cloud scenes. Specifically, Points2Sound consists of a vision network and an audio network. The vision network uses 3D sparse convolutions to extract a visual feature from the point cloud scene. Then, the visual feature conditions the audio network, which operates in the waveform domain, to synthesize the binaural version. Results show that 3D visual information can successfully guide multi-modal deep learning models for the task of binaural synthesis. We also investigate how 3D point cloud attributes, learning objectives, different reverberant conditions, and several types of mono mixture signals affect the binaural audio synthesis performance of Points2Sound for the different numbers of sound sources present in the scene.
\end{abstract}

\section{Introduction}
\label{sec:intro}

People perceive the world through multiple senses that jointly collaborate to understand the environment. While visual stimuli are important for spatial cognition, auditory stimuli are particularly critical. For example, being capable of hearing instantly from all angles helps people orient themselves in space and influences their visual attention\cite{culling2010spatial, robinson2013audition}. As auditory stimuli are received by both ears, our brain locates sound sources in space by comparing the sound that our ears receive. This process, known as binaural hearing, relies mainly on two acoustic cues: interaural time difference (ITD) and interaural level difference (ILD). ITD is the difference in the arrival time of a sound between the ears and ILD is the difference in sound intensity. In the median plane, i.e.\ the vertical plane between the ears, ITD and ILD are both small, and we rely on spectral cues to locate sources \cite{blauert1997spatial}. All such acoustic cues can be described by the head-related transfer function (HRTF), which encodes the sound distortion caused by geometries of the head and the torso~\cite{shaw1982external}.

In immersive applications, the generation of accurate binaural acoustic cues that match the visual counterpart is key to providing people with meaningful experiences in the virtual environment. These acoustic cues, ITD and ILD, strongly rely on the 3D position between the receiver and the sound sources. Recently, several methods using neural networks have been proposed for generating binaural audio from mono audio, using 2D visual information as guidance~\cite{zhou2020sep, yang2020telling, gao20192, lu2019self}. However, using 2D visual information inherently restricts the neural network's ability to extract information about the 3D positions between the receiver and the sound sources present in a scene. Not having access to this potentially useful information entails the risk of finding a sub-optimal solution for the task of binaural generation. In addition, these recent methods extract visual information by applying 2D dense convolutions to planar projections of the scene. This process forces the 2D convolutional filters to attend to local planar-projection regions with no relationship to physical space — possibly hindering the audio-visual learning necessary for the binaural synthesis task.

\begin{figure}[t]
    \center
    \includegraphics[scale=0.9]{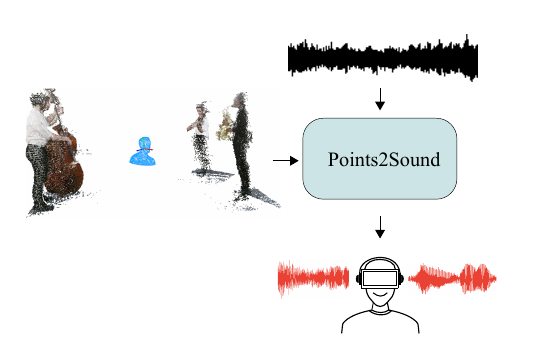}
    \caption{Points2Sound is a deep-learning model capable of generating a binaural version from mono audio that matches a 3D point cloud scene.}
    \label{fig:teaser}
\end{figure}

In this paper we introduce Points2Sound, a multimodal neural network that synthesizes binaural audio from mono audio using 3D visual information as guidance (see Fig.~\ref{fig:teaser}). For the visual learning, we propose the use of 3D point clouds as visual information as well as 3D sparse convolutions for extracting information from the 3D point clouds. This approach enables the model to extract information about the 3D position of the sound sources while convolutional filters attend to 3D data structures in local regions of the 3D space. For the audio learning, Points2Sound uses advancements in neural audio modeling in the waveform domain. We extend the Demucs architecture\cite{defossez2019music} and show that this model can be effectively conditioned by using visual information. 
Although Demucs was originally designed for source separation, we find it appropriate for binaural synthesis given that the model needs to intrinsically separate the sound of the sources in the mono mixture for further binaural rendering. This study thus analyzes the performance of Points2Sound for different 3D point cloud attributes, learning objectives, reverberant conditions and types of mono mixtures.

The main contributions of this work are the following:
\begin{itemize}

\item We introduce the use of 3D point clouds to condition audio signals. By using 3D visual information and 3D sparse convolutions, the neural network can learn the correspondence between audio characteristics (e.g. spatial cues or timbre) and 3D structures found in local regions of the 3D space — a correspondence relevant for binaural audio synthesis.

\item We tackle visually-informed binaural audio generation directly in the waveform domain, thereby optimizing our model in an end-to-end fashion and allowing it to learn audio features that are not limited by a fixed resolution of a spectrogram representation.

\item We evaluate how 3D point cloud attributes, i.e. depth or rgb-depth, several types of mono mixture signals, the effect of the room, and different learning objectives affect binaural audio synthesis performance for different numbers of sound sources in the scene.

\item We provide a dataset of the captured 3D-video point clouds, videos with listening examples, and the source code for reproducibility purposes\footnote{https://github.com/francesclluis/points2sound}.

\end{itemize}

This paper is organized as follows: Section \ref{sec:related_work} provides a brief overview of the related work. Section \ref{sec:approach} details the neural network architecture, the training procedure, and the data used. Section \ref{sec:results} presents the evaluation metrics and the obtained results. Section \ref{sec:discussion} discusses the results, and Section \ref{sec:conclusion} concludes.

\section{Related work}
\label{sec:related_work}

We provide a brief overview of related works in the field of audio-visual source separation and audio-visual spatial audio generation.

\subsection{Audio-visual Source Separation:} Source separation has been traditionally approached using only audio signals \cite{cardoso1998blind, haykin2005cocktail} with methods such as independent component analysis \cite{hyvarinen2000independent}, sparse coding \cite{olshausen1997sparse} or non-negative matrix factorization \cite{lee2000algorithms}. Recently, audio source separation has experienced significant progress due to the application of deep learning methods \cite{zeghidour2021wavesplit, luo2019conv, samuel2020meta, takahashi2021densely}. Current trends include performing source separation in the waveform domain \cite{stoller2018wave, lluis2018end, defossez2019music} or preserving binaural cues during the separation process \cite{han2020real, tan2020sagrnn}. 
In addition, deep learning methods have facilitated the inclusion of visual information to guide the audio separation \cite{ephrat2018looking, owens2018audio}. In the case of music source separation using visual information, learning methods mainly use appearance cues from the 2D visual representations \cite{zhao2018sound, gao2018learning}, but have been enhanced also with motion information \cite{zhao2019sound, gan2020music}. Interestingly, audio-visual source separation models intrinsically learn to map audio to their corresponding position in the visual representation. This has encouraged the use of visual information for spatial audio generation \cite{zhou2020sep}.

\subsection{Audio-visual Spatial Audio Generation:}

With the recent advances on audio modeling using neural networks, end-to-end deep learning approaches using explicit information about the position and orientation of the sources in the 3D space have been proposed for binaural audio synthesis \cite{richard2020neural, gebru2021implicit}. These approaches require head tracking equipment to know the pose of the receiver and the sound source in the environment.
Concurrently, audio-visual learning for spatial audio generation has gained interest. Several methods have been proposed to infer spatial acoustic cues to mono audio from leveraging visual information. Morgado et al.~\cite{morgado2018self} proposed a learning method to generate the spatial version of mono audio guided by 360\degree~videos. Their approach is to predict the spatial audio in ambisonics format which can be later decoded as binaural audio for reproduction through headphones. Gao et al.~\cite{gao20192} show that directly predicting the binaural audio creates better 3D sound sensations. They propose a U-Net-like framework for mono-to-binaural conversion using normal field of view videos.
Since then, binauralization models using 2D visual information have been enhanced using different approaches such as using an auxiliary classifier \cite{lu2019self} or integrating the source separation task in the overall binaural generation framework \cite{zhou2020sep}. In addition, it has been shown that features from pretrained models on audio-visual spatial alignment tasks are beneficial for audio binauralization \cite{yang2020telling}. Note that many of these approaches use audio spectrogram representations while considering mono mixture signals represented as the sum of the two spatial channels in order to train their networks\cite{zhou2020sep, lu2019self, gao20192, yang2020telling}. It remains unclear how operating in the waveform domain and using other mono representations may affect the binaural synthesis performance.

\section{Approach}
\label{sec:approach}

We propose Points2Sound, a deep learning algorithm capable of generating a binaural version from a mono audio using the 3D point cloud scene of the sound sources in the environment.

\subsection{Problem Formulation}

Consider an audio mono signal $s_{m}\in\mathbb{R}^{1\times T}$ of length $T$ samples generated by $N$ sources $s_i$, with 
\begin{equation}
s_{m}(t)=\sum_{i=1}^{N}s_i(t)
\label{eq:mono}
\end{equation}
along with the 3D scene of the sound sources in the 3D space represented by a set of $I$ points $\mathcal{P}=\{\mathcal{P}_i\}_{i=1}^I$, and the corresponding binaural signal $s_{b}\in\mathbb{R}^{2\times T}$. We aim at finding a model $f$ with the structure of a neural network such that $s_b(t)=f(s_{m}(t), \mathcal{P})$. The binaural version $s_b(t)$ generated by $N$ sources $s_i$ is defined as: 

\begin{equation}
s_{b}^{L,R}(t)=\sum_{i=1}^{N}s_i(t)\circledast \mathrm{HRTF}(\varphi_i, \theta_i, \dis_i)|_{\mathrm{left,right}}
\label{eq:binaural}
\end{equation}

\noindent where $\mathrm{HRTF}(\varphi_i, \theta_i, \dis_i)|_{\mathrm{left,right}}$ is the head-related transfer function of the sound incidence at the specified $i$-source orientation $(\varphi_i, \theta_i)$ and distance $\dis_i$  for both left ($L$) and right ($R$) ears. Throughout this work, orientation and distance are defined based on a head-related coordinate system, i.e. the center of coordinates is considered the head of the listener. During the training of the model we consider generic HRTFs measured in an anechoic environment. However during testing, we also evaluate the model performance under reverberant room conditions by using two binaural room impulse responses. Also note that in this work only the sound source’s contribution through the 3D point cloud is explicitly considered for binauralization. The contributions of the listener and the environment are implicitly considered via the choice of HRTF.

\subsection{Points2Sound Model Architecture}

\begin{figure*}[t]
    \center
    \includegraphics[scale=0.8]{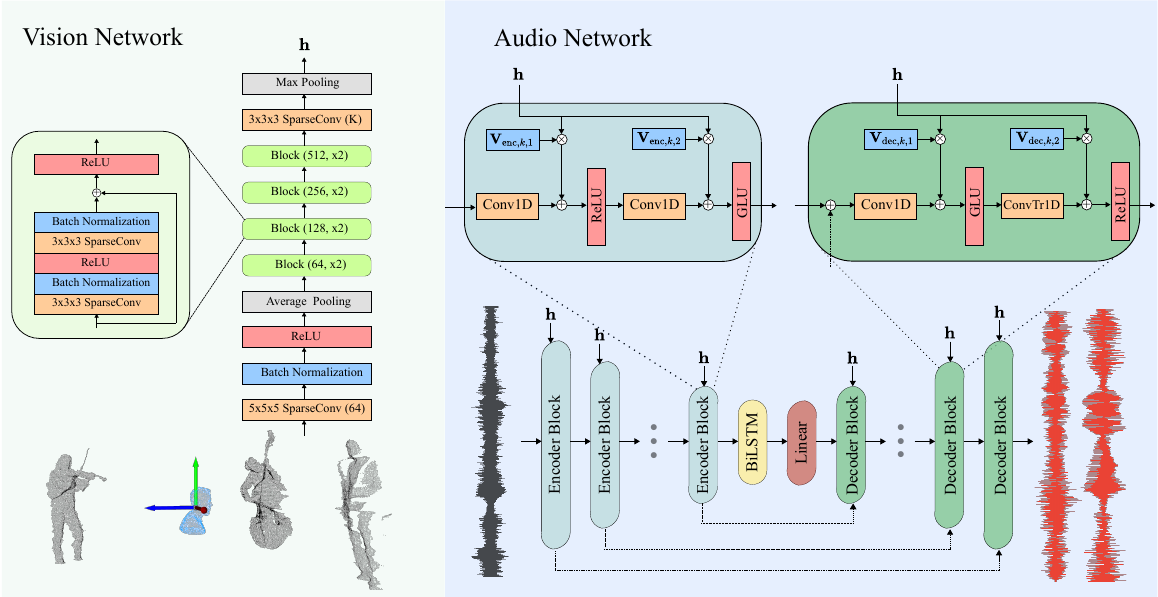}
    \caption{Overview diagram of Points2Sound. It consists of a sparse Resnet18 network for visual analysis and a Demucs network for binaural audio synthesis. The vision network extracts a visual feature $\mathbf{h}$ from the 3D point cloud. Then, this visual feature serves to condition the audio network to generate a binaural version from the mono audio that matches the visual counterpart. Both networks are jointly optimized during the training of the model.}
    \label{fig:main_diagram}
\end{figure*}

We propose a multi-modal neural network architecture capable of synthesizing binaural audio in an end-to-end fashion: the architecture takes as inputs the 3D point cloud scene along with the raw mono waveform and outputs the corresponding binaural waveform. The architecture comprises a vision network and an audio network. Broadly, the vision network extracts a visual feature from the 3D point cloud scene that serves to condition the audio network for binaural synthesis. Figure~\ref{fig:main_diagram} shows a schematic diagram of the proposed model.

\subsubsection{Sparse Tensor for 3D Point Cloud Representation}

3D scenes captured by rgb-depth cameras or LiDAR scanners can be represented by a 3D point cloud. A 3D point cloud is a low-level representation of a 3D scene that consists of a collection of $I$ points $\{\mathcal{P}_i\}_{i=1}^I$. The essential information associated to each point $\mathcal{P}_i$ is its location in space. For example, in a Cartesian coordinate system each point $\mathcal{P}_i$ is associated with a triple of coordinates $\mathbf{c}_i = (x_i, y_i, z_i)\in\mathbb{R}^{3}$ in the x,y,z-axis. In addition, each point can have associated features $\mathbf{f}_i\in\mathbb{R}^{n}$ like its color.

Extracting information from a 3D point cloud using neural networks requires non-standard operations that can handle the 3D data sparsity. It is common to represent the 3D point cloud information using a sparse tensor \cite{choy2020high, xie2020pointcontrast, gwak2020generative} and define operations on that sparse tensor, such the 3D sparse convolution operation. Note that sparse tensors require a discretization step that enables point cloud coordinates to be defined in the integer grid of the sparse tensor. 
In this work, the 3D point cloud is represented with a third order tensor by first discretizing its coordinates using a voxel size $v_s$. The voxel size $v_s$ denotes the discretization step size, and allows to define point cloud coordinates in the integer grid of the tensor. The discretized coordinates of each point are given by $\mathbf{c}_i^\prime = \lfloor\frac{\mathbf{c}_i}{v_s}\rfloor = (\lfloor\frac{x_i}{v_s}\rfloor, \lfloor\frac{y_i}{v_s}\rfloor, \lfloor\frac{z_i}{v_s}\rfloor)$. Then, the resultant tensor representing the point cloud is given by 

\begin{equation}
\mathbf{T}[\mathbf{c}_i^\prime]=
\left\{
	\begin{array}{ll}
		\mathbf{f}_i  & \mbox{if } \mathbf{c}_i^\prime\in\textit{C}^\prime\\
		0 & \mbox{otherwise},
	\end{array}
\right.
\end{equation}
where $\textit{C}^\prime$ is the set of discretized coordinates of the point cloud and $\mathbf{f}_i$ is the feature associated to the point $\mathcal{P}_i$.  
Note that in the following we will evaluate how 3D point cloud attributes affect the binaural audio synthesis. Accordingly, we will consider two types of 3D point cloud scenes: when 3D point cloud scenes consist of depth-only data and when 3D point cloud scenes consist of rgb-depth data. In the cases where depth-only information is available, we use the non-discretized coordinates as the feature vectors associated to each point, i.e. $\mathbf{f}_i=\mathbf{c}_i$. When rgb-depth information is available, we use the rgb values as the feature vectors associated to each point, i.e. $\mathbf{f}_i=(\text{r}_i, \text{g}_i, \text{b}_i)$.

\subsubsection{3D Sparse Convolution on a 3D Sparse Tensor}

The 3D sparse convolution is a generalized version of the conventional dense convolution designed to operate on a 3D sparse tensor\cite{choy20194d}. 

The 3D sparse convolution on a 3D sparse tensor is defined as follows:
\begin{equation}
\mathbf{T}^{\text{out}}[x, y, z]= \sum_{p,j,k\in\mathcal{N}(x,y,z)} \mathbf{W}[p,j,k]\mathbf{T}^{\text{in}}[x+p,y+j,z+k]
\end{equation}
for $(x,y,z)\in \textit{C}^\prime_\mathrm{out}$. Where  $\mathcal{N}(x,y,z)=\{(p,j,k)||p|\leq \Gamma, |j|\leq \Gamma, |k|\leq \Gamma, (p+x,j+y,k+z)\in \textit{C}^\prime_\mathrm{in}\}$. $\mathbf{W}$ are the weights of the 3D convolutional kernel and $2\Gamma+1$ is the convolution kernel size. $\textit{C}^\prime_\mathrm{in}$ and $\textit{C}^\prime_\mathrm{out}$ are predefined input and output discretized coordinates of sparse tensors\cite{choy20194d}.

\subsubsection{Vision Network} 
The vision network consists of a Resnet18~\cite{he2016deep} architecture with 3D sparse convolutions~\cite{choy20194d} that extracts a visual feature from the 3D point cloud scene. Resnet18 with 3D sparse convolutions has been successfully used in several tasks such as 3D semantic segmentation~\cite{choy20194d} or 3D single-shot object detection~\cite{gwak2020generative}.
Thus, we consider sparse Resnet18 suitable for our scenario, where extracting information about the position of the sources while recognizing the type of source is critical for reliable binaural synthesis.
Sparse Resnet18 learns at different scales by halving the feature space after two residual blocks and doubling the receptive field by using a stride of 2. A key characteristic of residual blocks are their residual connections which allow to propagate the input data through later parts of the block by skipping some layers. Through the network, ReLU is used as activation function and batch normalization is applied after sparse convolutions. At the top of Resnet18 4$^{th}$ block, we add a 3x3x3 sparse convolution with $K=16$ output channels and apply a max-pooling operation to adequate the dimensions of the extracted visual feature $\mathbf{h}$.

\subsubsection{Audio Network} 
We adapt the Demucs architecture~\cite{defossez2019music} to synthesize binaural versions $\hat{s}_b$ from mono audio signals $s_m$ given the 3D scene of the sound sources $\mathcal{P}$. Although Demucs was originally designed for source separation, we find it appropriate for binaural synthesis because the model needs to intrinsically learn to separate the sound of the sources in the mono mixture for further rendering (see Eq.~\ref{eq:binaural}). Demucs works in the waveform domain and has a U-Net-like structure~\cite{ronneberger2015u} (see Fig.~\ref{fig:main_diagram}). The encoder-decoder structure learns multi-resolution features from the raw waveform while skip connections allow low-level information to be propagated through the network. In the current case, skip connections allow later decoder blocks of the network to access information related to the phase of the input signal, which otherwise may be lost when propagated through the network. In this work, we keep the original six convolution blocks for both encoder and decoder but extend the architecture so that the input and output channels match our mono and binaural signals.

\subsubsection{Conditioning} 
We use a global conditioning approach on the audio network to guide the binaural synthesis according to the 3D scene. Global conditioning was introduced in Wavenet\cite{oord2016wavenet} and has been recently used in the Demucs architecture for separation purposes using one-hot vectors \cite{jenrungrot2020cone}. In a similar way, we use the extracted visual feature $\mathbf{h}$ from the vision network and insert it in each encoder and decoder block of the audio network. Specifically, the visual feature is inserted after being multiplied by a learnable linear projection $\mathbf{V}_{\cdot, q, \cdot}$. As in \cite{jenrungrot2020cone}, Demucs encoder and decoder takes the following expression:

\begin{align}
    \phantom{\texttt{Encode}}
    &\begin{aligned}
        \mathllap{\texttt{Encoder}_{q+1}} &= \text{GLU}(
        \mathbf{W}_{\text{encoder},q,2} \ast \text{ReLU}(\mathbf{W}_{\text{encoder},q,1} \ast \\  
        &\texttt{Encoder}_{q} + \mathbf{V}_{\text{encoder},q,1} \mathbf{h}) + \mathbf{V}_{\text{encoder},q,2}  \mathbf{h}
        ),
    \end{aligned}\\
    &\begin{aligned}
        \mathllap{\texttt{Decoder}_{q-1}} &= \text{ReLU}(
        \mathbf{W}_{\text{decoder},q,2} \ast^\top \text{GLU}(\mathbf{W}_{\text{decoder},q,1} \ast \\ & (\texttt{Encoder}_{q} + \texttt{Decoder}_{q}) 
         + \mathbf{V}_{\text{decoder},q,1} \mathbf{h})
         \\ & + \mathbf{V}_{\text{decoder},q,2} \mathbf{h}
        ).
    \end{aligned}
\end{align}
where $\texttt{Encoder}_{q+1}$ and $\texttt{Decoder}_{q-1}$ are the outputs from the $q\text{-th}$ level encoder and decoder blocks respectively. $\mathbf{W}_{\cdot, q, \cdot}$ are the 1-D kernel weights at the $q\text{-th}$ block. Rectified Linear Unit (ReLU) and Gated Linear Unit\cite{dauphin2017language} (GLU) are the corresponding activation functions. The operator $\ast$ denotes the 1-D convolution while $\ast^\top$ corresponds to a transposed convolution operation, as commonly defined in the deep learning frameworks \cite{paszke2019pytorch}.

\subsubsection{Learning Objective}
During the training of Points2Sound, the parameters of both vision and audio networks are optimized to reduce the L1 loss function between the estimated binaural signal $\hat{s}_b^{L,R}$ and the ground truth binaural signal $s_b^{L,R}$. The L1 loss computes the absolute error between the estimated and the ground truth waveform samples. We refer to this learning objective as 
\begin{equation}
\mathcal{L}_{\mathrm{full}} = \lVert s_{b}^{L,R} - \hat{s_{b}}^{L,R}\lVert.
\end{equation}

Note that in Section \ref{sec:evaluation} we will investigate the effect of another learning objective on the performance of Points2Sound.

\subsection{Data}
While there are lots of audio datasets, there is data scarcity of 3D point cloud videos of performing musicians. For the purposes of this work, we capture 3D videos of the same twelve performers playing different instruments: cello, doublebass, guitar, saxophone, and violin. In addition, we separately collect audio recordings of these instruments from existing audio datasets. This data will serve later to generate 3D audio-visual scenes for supervised learning.

\subsubsection{Point Clouds} 
Recordings were conducted using an Azure Kinect DK (by Microsoft) placed one meter above the floor and capturing a frontal view of the musician at a distance of two meters. Azure Kinect DK comprises a depth camera and a color camera. The depth camera was capturing a $75\degree \times 65\degree$ field of view with a $640\times 576$ resolution while the color camera was capturing with a $1920\times 1080$ resolution. Both cameras were recording at 15 fps and Open3D library\cite{Zhou2018} was then used to align depth and color streams and generate a point cloud for each frame. The full 3D video recordings span 1 hour of duration with an average of 12 performers for each instrument.

We increase our 3D point cloud video dataset collecting 3D videos from \textit{small ensemble 3D-video database} \cite{thery2019anechoic} and \textit{Panoptic Studio} \cite{joo2017panoptic}. In \textit{small ensemble 3D-video database}, recordings are carried out using three RGB-Depth Kinect v2 sensors. LiveScan3D \cite{kowalski2015livescan3d} and OpenCV libraries are then used to align and generate point clouds for each frame given each camera point of view and sensor data. The average video recording is 5 minutes per instrument and a single performer per instrument. In \textit{Panoptic Studio}, recordings are carried out using ten Kinect sensors. In this case, recordings span two instrument categories: cello and guitar. The average time per video recording is 2 minutes per instrument for a single performer per instrument.
As we gather 3D point cloud videos from different sources, we set the axes representing the point clouds to have the same meaning for all the collected 3D videos. This is: the body face direction is z-axis, stature direction is y-axis, and the side direction is x-axis. Then we split 75\% of the data for training, 15\% for validation, and the remaining 10\% for testing. Data split is made ensuring that there is no overlap in identities between sets.

\subsubsection{\label{subsec:audio}Audio} 
We collect 30 hours of mono audio recordings at 44.1~kHz from \textit{Solos} \cite{montesinos2020solos} and \textit{Music} \cite{zhao2018sound}. Both datasets gather music from YouTube which ensures a variety of acoustic conditions. In total, we gather 72 recordings per instrument with an average of 5 minutes per recording. We split 75\% of the recordings for training, 15\% for validation, and the remaining 10\% for testing.

For further binaural auditory scene generation, we also create multiple binaural versions of each recording using the \textit{Two!Ears} Binaural Simulator \cite{winter2017two}. Specifically, for each audio recording we simulate the binaural version at a discrete set of angular positions in the horizontal plane with no elevation, i.e.\ $(\varphi_k, \theta):= (\frac{k\pi}{4} ,0)$ for $k=0,\ldots,7$. For binaural auditory modeling, we use the HRTFs at 1m of distance between source and receiver measured with a KEMAR manikin (type 45BA) at the anechoic chamber of the TU Berlin \cite{wierstorf2011free}.

\subsection{\label{sec:scene-gen}Audio-visual 3D Scene Generation}

We synthetically create mono mixtures, 3D scenes, and the corresponding binaural version to train the model in a supervised fashion.

\begin{figure}[ht!]
    \center
    \includegraphics[scale=0.9]{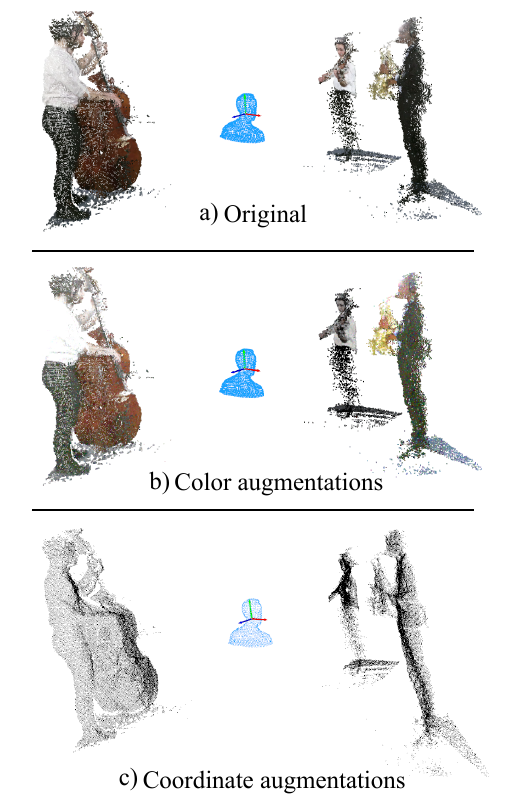}
    \caption{Illustration of the augmentation operations applied to a 3D point cloud scene. a) Original scene. b) Color augmentations applied to the original scene. Color augmentations include modifying the brightness and intensity of the sound sources as well as distort the color of each point using Gaussian noise. c) Coordinate augmentations applied to the original scene. Coordinate augmentations include random shearing along all axes and random translation along the stature direction.}
    \label{fig:augmentations}
\end{figure}

For each instance, we randomly select $N$ sources and $N$ angular positions with $N$ chosen uniformly between 1 and 3. The binaural mixture, which serves as supervision, is created following Eq.~\ref{eq:binaural}. First, we select 3 seconds length binaural signals for each sound source in the mix based on its angular position, and then we sum all selected binaural signals to create the binaural mixture.

For the 3D scene, we first select individual musician's point clouds corresponding to these sources. Then, musician point clouds are located at their corresponding angular position in a random distance ranging from 1 to 3 meters from the listeners head in the 3D space. Finally, all N musician point clouds are merged to create a single 3D point cloud scene. Note that we generate binaural versions using HRTFs computed at 1 meter distance from the listeners head but locate the sources in a distance ranging from 1 to 3 meters. This assumption is based on the fact that distance has a smaller influence on the shape of HRTFs, for source-receiver distances greater than 1m~\cite{otani2009numerical}.

During the training of the model, each individual musician point cloud is independently augmented in both coordinates and color. We randomly shear and translate the coordinates of each musician in the scene. Shearing is applied along all axes and the shear elements are sampled from a Normal distribution $\mathcal{N}(0, 0.1^2)$. Translation is applied along the stature direction and the translation offset is sampled from  $\mathcal{N}(0, 0.2^2)$.

Regarding color, we distort the brightness and intensity of each sound source in the scene. Specifically, we apply color distortion to each point via adding Gaussian noise sampled from $\mathcal{N}(0, 0.05^2)$ on each rgb color channel. We also alter color value and color saturation with random amounts uniformly sampled ranging from -0.2 to 0.2 and -0.15 to 0.15 respectively. Figure~\ref{fig:augmentations} illustrates the different augmentation operations applied. After augmentation, each scene is represented as a sparse tensor by discretizing the point cloud coordinates using a voxel size of 0.02 meters. We select a small voxel size as it has been shown to work better than bigger ones for several 3D visual tasks\cite{choy20194d}.

\subsection{Implementation Details}
Initially, we pretrain the vision network to facilitate the future learning process. Pretraining is done on the 3D object classification task modelnet40~\cite{wu20153d}. Since modelnet40 consists of 3D CAD models, we sample point clouds from the mesh surface of the objects shapes.  For the pretraining, we also discretize the coordinates setting the voxel size to 0.02 meters.
Then, Points2Sound vision and audio networks are jointly trained for 120k iterations using the Adam \cite{kingma2015adam} optimizer. We use a batch size of 40 samples and we set the learning rate to $1\times 10^{-4}$. We select the weights corresponding to the lowest validation loss after the training process. Training and testing are conducted on a single Titan RTX GPU. The training stage takes about 72 hours while the inference takes 0.115 seconds to binauralize 10 seconds of mono audio (value averaged from 300 samples). We use the Minkowski Engine \cite{choy20194d} for the sparse tensor operations and PyTorch \cite{paszke2019pytorch} for the other operations required.

\section{Results}
\label{sec:results}

\begin{table*}[ht!]
\center
\caption{Quantitative results of baselines and Points2Sound considering different mono input audio. For each method, we use rgb-depth 3D point cloud attributes and report the performance depending on the number of sources ($N = 1,2,3$), based on the average of the evaluation metrics. Average values for any number of sources are given by $\overline{\dis_\mathrm{ENV}}$ and $\overline{\dis_\mathrm{STFT}}$.}
\subcaption*{$s_{m}$ (True Mono)}
\begin{tabular}{l|ccc:ccc|c:c}
  \hline
   & \multicolumn{3}{c:}{$\dis_\mathrm{ENV} \downarrow$} & \multicolumn{3}{c|}{$\dis_\mathrm{STFT} \downarrow$} & $\overline{\dis_\mathrm{ENV}} \downarrow$ & $\overline{\dis_\mathrm{STFT}} \downarrow$ \\
   & 1 & 2 & 3 & 1 & 2 & 3 & & \\
 \hline
 Mono-Mono & 0.387 & 0.403 & 0.388 & 26.719 & 26.414 & 26.747 & 0.392 & 26.626 \\
 \hline
 Rotated-Visual & 0.232 & 0.285 & 0.305 & 9.002 & 10.588 & 12.016 & 0.274 & 10.535 \\
 \hline
 Points2Sound ($\mathcal{L}_{\mathrm{full}}$) & \textbf{0.173} & \textbf{0.248} & \textbf{0.280} & \textbf{3.297} & \textbf{6.645} & \textbf{9.080} & \textbf{0.233} & \textbf{6.340} \\
 \hline
\end{tabular}
\bigskip
\subcaption*{$s_{m} = s_{b}^L$}
\begin{tabular}{l|ccc:ccc|c:c}
  \hline
   & \multicolumn{3}{c:}{$\dis_\mathrm{ENV} \downarrow$} & \multicolumn{3}{c|}{$\dis_\mathrm{STFT} \downarrow$} & $\overline{\dis_\mathrm{ENV}} \downarrow$ & $\overline{\dis_\mathrm{STFT}} \downarrow$ \\
   & 1 & 2 & 3 & 1 & 2 & 3 & & \\
 \hline
 Mono-Mono & 0.148 & 0.155 & 0.159 & 7.472 & 6.997 & 6.951 & 0.154 & 7.14 \\
 \hline
 Rotated-Visual  & 0.165 & 0.166 & 0.165 & 7.610 & 6.808 & 6.345 & 0.165 & 6.921 \\
 \hline
 Points2Sound ($\mathcal{L}_{\mathrm{full}}$) & \textbf{0.054} & \textbf{0.103} & \textbf{0.130} &  \textbf{0.636} & \textbf{1.820} & \textbf{2.604} & \textbf{0.095} & \textbf{1.686} \\
 \hline
\end{tabular}
\bigskip
\subcaption*{$s_{m} = s_{b}^L+s_{b}^R$}
\begin{tabular}{l|ccc:ccc|c:c}
  \hline
   & \multicolumn{3}{c:}{$\dis_\mathrm{ENV} \downarrow$} & \multicolumn{3}{c|}{$\dis_\mathrm{STFT} \downarrow$} & $\overline{\dis_\mathrm{ENV}} \downarrow$ & $\overline{\dis_\mathrm{STFT}} \downarrow$ \\
   & 1 & 2 & 3 & 1 & 2 & 3 & & \\
 \hline
 Mono-Mono & 0.142 & 0.166 & 0.178 & 4.046 & 4.112 & 4.058 & 0.162 & 4.072 \\
 \hline
 Rotated-Visual  & 0.166 & 0.192 & 0.209 & 5.663 & 5.918 & 6.031 & 0.189 & 5.870 \\
 \hline
 Points2Sound ($\mathcal{L}_{\mathrm{full}}$) & \textbf{0.015} & \textbf{0.073} & \textbf{0.114} & 0.099 & \textbf{0.762} & \textbf{1.521} & \textbf{0.067} & \textbf{0.794} \\
 \hdashline
 Points2Sound ($\mathcal{L}_{\mathrm{full}}$) (only-depth) & 0.016 & 0.080 & 0.122 & \textbf{0.082} & 0.885 & 1.736 & 0.072 & 0.901 \\
 \hdashline
 Points2Sound ($\mathcal{L}_{\mathrm{diff}}$) & 0.015 & 0.090 & 0.125 & 0.153 & 1.205 & 1.832 & 0.076  & 1.063 \\
 \hline
\end{tabular}
\bigskip
\label{table:quantitative_input_mono_results}
\end{table*}

\subsection{Evaluation Metrics}
As in previous work\cite{morgado2018self}, we measure the quality of the predicted binaural audio assessing the short-time Fourier transform (STFT) Distance. Using the STFT Distance we assess how similar the frequency components of each predicted binaural channel are to the ground truth. STFT Distance ($\dis_\mathrm{STFT}$) between a binaural signal $s_{b}$ and its estimate $\hat{s_{b}}$ is defined as:

\begin{equation}
\begin{split}
\dis_\mathrm{STFT} = & \lVert \text{STFT}(s_{b}^{L}(t)) - \text{STFT}(\hat{s}_{b}^{L}(t)) \rVert_2\ + \\ &\lVert \text{STFT}(s_{b}^{R}(t)) - \text{STFT}(\hat{s}_{b}^{R}(t)) \rVert_2
\end{split}
\end{equation}
where $\lVert\cdot\rVert_2$ is the L2 norm and STFT($\cdot$) is the short-time Fourier transform. The STFT is computed using a Hann window of 23 milliseconds and hop length of 10 milliseconds.

We also assess the quality of the predicted binaural audio using the Envelope Distance. The Envelope Distance operates in the time domain and is intended to capture the perceptual similarity between two binaural signals in a better way than directly computing the loss between its waveform samples\cite{morgado2018self}. Envelope Distance ($\dis_\mathrm{ENV}$) between a binaural signal $s_{b}$ and its estimate $\hat{s_{b}}$ is defined as:

\begin{equation}
\begin{split}
\dis_\mathrm{ENV} = & \lVert E[s_{b}^{L}(t)] - E[\hat{s}_{b}^{L}(t)]\rVert_2\ + \\ &\lVert E[s_{b}^{R}(t)] - E[\hat{s}_{b}^{R}(t)]\rVert_2
\end{split}
\end{equation}
where $E[s(t)]$ corresponds to the envelope of the signal $s(t)$. The envelope is given by the magnitude of the analytical signal computed using the Hilbert transform.

Note that we report the performance depending on the number of sources ($N = 1,2,3$), based on the average of the evaluation metrics. Average values for any number of sources are given by $\overline{\dis_\mathrm{ENV}}$ and $\overline{\dis_\mathrm{STFT}}$. Also, before computing the distances, the predicted and the ground truth signals are normalized according to their maximum absolute value.

\subsection{Baselines}

We use two baselines to assess the quality of the predicted binaural versions:

\paragraph{\textbf{Rotated-Visual.}} Rotated-Visual baseline assesses the performance of Points2Sound ($\mathcal{L}_{\mathrm{full}}$) when wrong visual information is provided. To this end, during testing we rotate the 3D scene by $\pi/2$ in the horizontal plane of the listener's head. 

\paragraph{\textbf{Mono-Mono.}} Mono-Mono baseline simply copies the mono input audio to both binaural predicted channels.

\paragraph{}A quantitative comparison with a similar method for mono to binaural synthesis using visual information is provided in the Appendix.

\subsection{Evaluation}
\label{sec:evaluation}

For evaluation we use 504 audio-visual 3D scenes with $N=1, 2, 3$ sound sources. Audio-visual 3D scenes are generated using the test data set and following the procedure explained in Section~\ref{sec:scene-gen}. During evaluation, augmentation operations are not applied and 10 second audio clips are selected.

\textbf{Points2Sound input audio}.

We consider three different types of mono mixture input signals for Points2Sound. Table \ref{table:quantitative_input_mono_results} shows quantitative results of Point2Sound for the different types of mono mixture signals and number of sources. 

First, we consider true mono mixture signals which come from the audio dataset, detailed in \ref{subsec:audio}, where no HRTFs have been applied. In this case, Points2Sound improvement over the baselines is notable. This is especially observed in the $\overline{\dis_\mathrm{STFT}}$ metric, where the Mono-Mono baseline achieves 26.626 while Points2Sound achieves 6.340. Despite the improvement, the binaural predictions are degraded and contain time-frequency artifacts that change the timbre characteristics of the original sound.

Second, we consider mono mixture signals which contain only the left binaural channel, i.e. $s_{m} = s_{b}^L$. Note that in this case, ITDs are not preserved in the mono mixture and the network has to shift differently between left and right channels in the binaural prediction. Results show that Points2Sound improves the evaluation metrics for all baselines especially when few sources are present. When $N=1$ source, Points2Sound achieves a $\dis_\mathrm{ENV}$ of 0.054 and a $\dis_\mathrm{STFT}$ of 0.636, as opposed to 0.165 and 7.610 achieved by the Rotated-Visual baseline. 
It is important to remark that in comparison to the true mono input signal, the absolute values in $\dis_\mathrm{STFT}$ show how much information is given to the model when the HRTFs are already applied in the mono mixture.

Third, following previous work \cite{zhou2020sep, lu2019self, gao20192, yang2020telling}, we consider mono mixture signals represented as the sum of the two spatial channels, i.e. $s_{m} = s_{b}^L+s_{b}^R$. In this case the mixing of the channels creates a mono signal that loses spatial properties. But, the resultant mono signal preserves the correct ITDs from the binaural version. Results show that Point2Sound achieves the best results using this mono representation with a $\overline{\dis_\mathrm{ENV}}$ of 0.095 and a $\overline{\dis_\mathrm{STFT}}$ of 1.686. Note that the obtained quantitative results are similar to the ones achieved using the above approach, where mono mixture signals are represented as $s_{m} = s_{b}^L$.

For the following, when we refer to Points2Sound we assume it has been trained using mono mixture signals represented as $s_{m} = s_{b}^L+s_{b}^R$.

\textbf{3D point cloud attributes.} We evaluate how 3D point cloud attributes affect the binaural audio synthesis. Accordingly, we consider two types of 3D point cloud scenes: when 3D point cloud scenes consist of depth-only data and when 3D point cloud scenes consist of rgb-depth data. We use the term ``only-depth'' in Table \ref{table:quantitative_input_mono_results} to report the performance when 3D point cloud scenes consist of depth-only data. Otherwise, we report the performance when 3D point cloud scenes consist of rgb-depth data.

We observe that Points2Sound benefits from the rgb information especially when multiple sound sources are present. For example with $N=3$ sources, Points2Sound using rgb-depth features achieves a $\dis_\mathrm{ENV}$ of 0.114 and a $\dis_\mathrm{STFT}$ of 1.521, as opposed to 0.122 and 1.736 achieved by only depth features. But with $N=1$ source, Points2Sound using depth features slightly outperform rgb-depth features, providing a $\dis_\mathrm{STFT}$ of 0.082 and a $\dis_\mathrm{ENV}$ of 0.016, as opposed to 0.099 and 0.015  We suspect that with a single source, depth features already provide straight information about the position of the source for accurate binaural synthesis. However, when multiple sources are present, rgb-depth features better distinguish each source which facilitates the binaural synthesis of the audio network. Informal listening corroborates that rgb-depth features helps the model to recognize and locate multiple sources, as it provides more stable auditory images of the sound sources for the whole 10 second clips.

\textbf{Number of sources in the 3D scene.} We are also interested in evaluating the quality of the predicted binaural audio depending on the number of sound sources present in the 3D scene. We observe that when $N=1$ sources, Points2Sound provides perceptually convincing binaural predictions with a consistent performance across almost all examples.

When $N=2$ sources, informal listening reveals that binaural audio predictions are convincing especially when the sources are located at the same side of the listener's head. Quantitative results show that Points2Sound using rgb-depth features achieves a $\dis_\mathrm{ENV}$ of 0.044 and a $\dis_\mathrm{STFT}$ of 0.284 when both sources are in the same side and a $\dis_\mathrm{ENV}$ of 0.084 and a $\dis_\mathrm{STFT}$ of 0.944 for other source position configurations. We also observe that in some cases, Points2Sound has difficulties when one of the two sources is located in front or behind the listener's head. This results in binaural predictions where the auditory image of the front/back source is not stable for the whole 10 seconds clip.

When N=3 sources, Points2Sound has difficulties to provide stable auditory images for every sound source for the whole 10 seconds clip.

\begin{figure}[ht!]
    \center
    \includegraphics[scale=0.9]{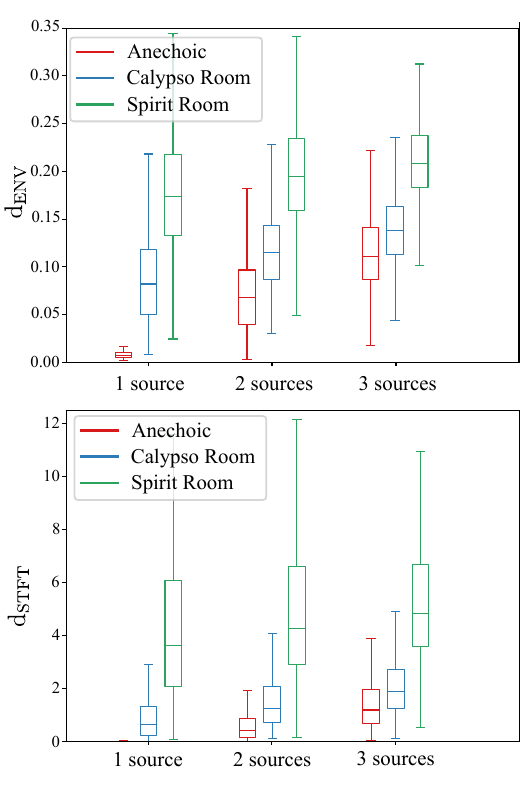}
    \caption{Envelope Distance ($\dis_\mathrm{ENV}$) and STFT Distance ($\dis_\mathrm{STFT}$) of Points2Sound ($\mathcal{L}_{\mathrm{full}}$), trained in anechoic conditions, for different rooms and number of sources in the 3D scene. The results are reported using rgb-depth point cloud features. For each number of sources, the box extension shows the first and third quartile of the data with a line at the median. The whiskers extending from the box show the range of the data.}
    \label{fig:box_plot_results_room}
\end{figure}

\textbf{Effect of the room.} We evaluate how Points2Sound performs under reverberant room conditions. To this end, we use binaural room impulse responses (BRIRs) to generate two test sets. Note that Points2Sound is trained using HRTFs measured in an anechoic room. 
The first test set is generated using the BRIRs measured in the studio room Calypso at TU Berlin\cite{wierstorf_hagen_2016_160761}. The Calypso room has a volume of 83 $\text{m}^3$ and a reverberation time RT60 of 0.17 seconds at a frequency of 1 kHz. 
The second test set is generated using the BRIRs measured in the meeting room Spirit at TU Berlin\cite{ma2015machine}. The Spirit room has a rectangular shape with an estimated reverberation time RT60 of 0.5 seconds. In both rooms, BRIRs  were measured using a KEMAR manikin (type 45BA) and a loudspeaker (Genelec 8250A) placed in front of the manikin at two meters of distance. 
As the head movements of the manikins were measured from -$\pi/2$ to $\pi/2$, we simulate the binaural versions at the following discrete set of angular positions in the horizontal plane with no elevation $(\varphi_k, \theta):= (\frac{k\pi}{4} ,0)$ for $k=-2,\ldots,2$. Finally, 504 audio-visual 3D scenes with $N=1, 2, 3$ are generated using the same procedure explained in Section~\ref{sec:scene-gen}.
Figure \ref{fig:box_plot_results_room} shows a visual comparison of the performance of Points2Sound for different room acoustic conditions. The results are reported for a different number of sources using rgb-depth point cloud features. 
We observe that Points2Sound performance decreases as the testing of acoustic conditions diverge from the anechoic training. In this case, better binaural predictions are achieved in the room Calypso with a RT60 of 0.17 seconds as opposed to predictions in the room Spirit with a RT60 of 0.5 seconds. It is in high reverberant rooms where it becomes evident that Points2Sound has not been trained to model room effects such as reflections, especially for low number of sources in the scene. For example with $N=1$ sources, Points2Sound achieves an average of $\dis_\mathrm{STFT}$ of 4.491 in the Spirit room in contrast with 1.081 achieved in the Calypso room. Points2Sound performance in the dry room, i.e. Calypso room, is closer to the anechoic performance, in particular when multiple sound sources are present. With $N=3$ sources, Points2Sound achieves an average of $\dis_\mathrm{STFT}$ 2.180 in Calypso room and 1.521 in the anechoic setting. This suggests that Points2Sound potential applicability should consider dry rooms that resemble the anechoic training conditions.

\textbf{Points2Sound loss function.} We analyze Points2Sound performance depending on the learning objective. The initial proposed learning objective, referred to as $\mathcal{L}_{\mathrm{full}}$, optimizes the parameters of the model to reduce the L1 loss between the estimated binaural $\hat{s_{b}}^{L,R}$ and the ground truth binaural $s_{b}^{L,R}$, i.e.
\begin{equation}
\mathcal{L}_{\mathrm{full}} = \lVert s_{b}^{L,R} - \hat{s_{b}}^{L,R}\lVert
\end{equation}
Note that in this case, Points2Sound predicts the full binaural signal, i.e. predicts both left and right binaural channels.
Several methods in the literature propose to optimize the models by predicting the difference of the two binaural channels\cite{gao20192, zhou2020sep}. To this end, we consider another loss function for Points2Sound, i.e.  $\mathcal{L}_{\mathrm{diff}}$, which optimizes the parameters to reduce the L1 loss between the estimated binaural difference channels $\hat{s_{b}}^{\mathrm{diff}}$ and the ground truth binaural difference channels $s_{b}^{\mathrm{diff}}$, i. e.
\begin{equation}
\mathcal{L}_{\mathrm{diff}} = \lVert s_{b}^{\mathrm{diff}} - \hat{s_{b}}^{\mathrm{diff}}\lVert
\end{equation}
where $s_{b}^{\mathrm{diff}}(t)= s_{b}^{L}-s_{b}^{R}$. 
Note that when using $\mathcal{L}_{\mathrm{diff}}$, Point2Sound is forced to learn the differences between the left and right binaural channels and predicts a one-channel signal $\hat{s_{b}}^{\mathrm{diff}}$. Then, considering the mono signal represented as $s_{m} = s_{b}^L+ s_{b}^R$, both predicted binaural channels are recovered as follows:
\begin{equation}
\hat{s_{b}}^L = (s_{m}+\hat{s_{b}}^{\mathrm{diff}})/2, \quad \hat{s_{b}}^R = (s_{m}-\hat{s_{b}}^{\mathrm{diff}})/2 
\end{equation}

Results in Table \ref{table:quantitative_input_mono_results} show that Points2Sound benefits from directly predicting the binaural signal using the $\mathcal{L}_{\mathrm{full}}$ loss function as opposed to predicting the difference between binaural channels with the $\mathcal{L}_{\mathrm{diff}}$ loss function. Using rgb-depth point cloud features, Points2Sound ($\mathcal{L}_{\mathrm{full}}$) achieves a $\overline{\dis_\mathrm{ENV}}$ of 0.067 and a $\overline{\dis_\mathrm{STFT}}$ of 0.794 while Points2Sound ($\mathcal{L}_{\mathrm{diff}}$) achieves 0.076 and 1.063 respectively. The poor performance obtained with the rotated-visual baseline indicates that Points2Sound strongly relies on the 3D scene to synthesize binaural audio and incorrect predictions are expected when using wrong visual information. In the following, we refer to Points2Sound assuming it has been trained using the $\mathcal{L}_{\mathrm{full}}$ loss function.

\subsection{\label{subsec:listening_examples}Listening Examples}

We provide a supplementary video with four listening examples where Points2Sound is applied to real-world data we record from expert musicians. We consider four challenging audio-visual scenes of $N=2$ sources performing simultaneously in the same room. Specifically, two audio-visual scenes contain guitar and violin as sound sources while the other two contain doublebass and violin. The recorded audio fragments cover a variety of music styles (classical and jazz), tempi (vivace and lento), and dynamics (forte and piano). The 3D scenes of musicians are captured using Azure Kinect DK cameras while mono audio is captured using a Google Pixel 4 smartphone at a static position in the middle of the room. For each scene, the video shows the raw data first, and then demonstrates the binaural predictions of Points2Sound. Despite the discrepancy between training data and real-world scenarios, the binaural predictions of Points2Sound show promising extrapolation ability.

\section{Discussion}
\label{sec:discussion}
The work presented in this paper indicates the potential of using 3D visual information to guide multi-modal deep learning models for the synthesis of binaural audio from mono audio. By using 3D point clouds as visual information, the vision network has the ability to extract information about the 3D positions between the receiver and the sound sources in a scene to guide the binaural synthesis. By using 3D sparse convolutions, the network learns the correspondence between 3D structures found in local regions of the 3D space and audio characteristics.

When Points2Sound is trained using true mono signals that do not contain HRTFs information, our proposed method introduces time-frequency artifacts that lead to degraded binaural predictions. This suggests that a significant amount of Points2Sound's capacity is needed to model the HRTFs information. As a result, the model has more difficulties to synthesize accurate binaural sound. Considering one of the two binaural channels as input mono, i.e. $s_{m} = s_{b}^L$, Points2Sound achieves similar quantitative results as when considering the mono audio as the sum of the two binaural channels, i.e $s_{m} = s_{b}^L+s_{b}^R$. Interestingly, the model trained using the sum of the two binaural channels as mono input provides encouraging extrapolation results when applied to real mono recordings, as demonstrated by the provided sound examples.

Results suggest that waveform-based approaches can provide convincing performance for the task of visually-informed spatial audio generation without the need to rely on hand-crafted spectrograms as input. In addition, by operating in the waveform domain, our model synthesizes the signal directly. This is in contrast to spectrogram-based models which predict a mask to overcome the difficulties of directly predicting the spectrum, due to the large dynamic range of STFTs\cite{zhou2020sep, lu2019self, gao20192, yang2020telling}.

Our proposed  model benefits from predicting the full binaural signal as opposed to the difference between binaural channels. This might be of relevance for other applications where visually-informed models operating in the waveform domain are used to generate spatial audio.

Our proposed model benefits from using visual features extracted from rgb-depth point clouds to improve the binaural synthesis when multiple sources are present, in comparison with features extracted from depth-only point clouds. 
However, the fact that Points2Sound can work with only-depth information may be beneficial in cases of low ambient light, where RGB sensors would fail to capture the scene, in contrast to LiDAR sensors that are still able to capture depth information.
As mentioned above, we observe that in some cases Points2Sound predicts binaural versions where the auditory image of the sources is not stable. As this effect is mainly observed for cases with the number of sources $N>1$, we suspect that this problem is related to the source separation capability of the audio network. To further investigate this phenomenon, a separate study on channel bleeding in source separation for different types of musical instruments would be required.

After analyzing the performance of Points2Sound under reverberant conditions, it is shown that the method could be applied to dry rooms that resemble the anechoic training conditions. However, a decreased performance is expected when the room acoustics conditions diverge from the anechoic training. The performance of Points2Sound in highly reverberant rooms, after retraining or fine-tuning the model using binaural room impulse responses that contain the influence of the room, remains to be studied.

\section{Conclusion and future work}
\label{sec:conclusion}

This work introduced Points2Sound, a multi-modal deep learning model capable of generating a binaural version from mono audio using a 3D point cloud scene as guidance. Points2Sound shows that 3D visual information can successfully guide the binaural synthesis while demonstrating that waveform-based approaches can provide convincing performance for the task of visually-informed spatial audio generation.

Such models see increased interest for the generation of spatial audio in immersive applications. Recent portable devices, like smartphones, have the ability to capture 3D visual data from the environment using LiDAR or rgb-depth cameras. However, such devices have limited capabilities to record spatial audio from the sound sources. Having a recorded rgb-depth environment and its corresponding mono audio, our approach is a step towards synthesizing proper acoustic stimuli for the users navigating the virtual environment depending on their location and head position. 

Future work could involve adding loudness into the learning process via predicting a reference sound level for each source. This would allow to infer also sound attenuation in 3D dynamic scenes.

\section*{Appendix}

\begin{figure}[ht!]
    \center
    \includegraphics[scale=0.9]{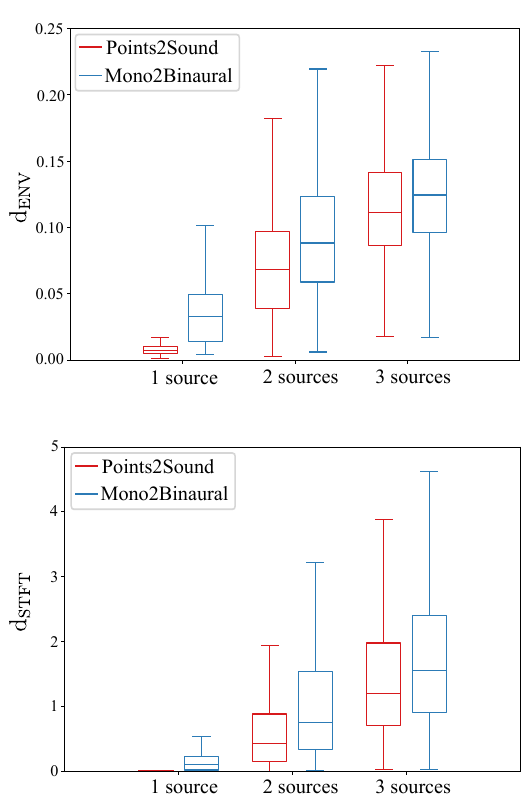}
    \caption{Envelope Distance ($\dis_\mathrm{ENV}$) and STFT Distance ($\dis_\mathrm{STFT}$) of Points2Sound using $\mathcal{L}_{\mathrm{full}}$ loss function and Mono2Binaural for different number of sources in the 3D scene. The results are reported using rgb-depth point cloud features. For each number of sources, the box extension shows the first and third quartile of the data with a line at the median. The whiskers extending from the box show the range of the data.}
    \label{fig:box_plot_results}
\end{figure}

\begin{table*}[ht!]
\center
\caption{Quantitative results of Points2Sound and Mono2Binaural. For each method, we report the performance depending on the number of sources ($N = 1,2,3$) and the type of 3D point cloud attributes (depth or rgb-depth), based on the average of the evaluation metrics. Average values for any number of sources are given by $\overline{\dis_\mathrm{ENV}}$ and $\overline{\dis_\mathrm{STFT}}$.}
\begin{tabular}{ll|ccc:ccc|c:c}
  \hline
  & Visual features & \multicolumn{3}{c:}{$\dis_\mathrm{ENV} \downarrow$} & \multicolumn{3}{c|}{$\dis_\mathrm{STFT} \downarrow$} & $\overline{\dis_\mathrm{ENV}} \downarrow$ & $\overline{\dis_\mathrm{STFT}} \downarrow$ \\
   & & 1 & 2 & 3 & 1 & 2 & 3 & & \\
 \hline
 \multirow{2}{*}{Mono2Binaural\cite{gao20192}} & depth & 0.038 & 0.101 & 0.132 & 0.192 & 1.300 & 2.029 & 0.090 & 1.174 \\
 & rgb-depth & 0.036 & 0.094 & 0.126 & 0.213 & 1.142 & 1.858 & 0.085 & 1.071 \\\hline
 \multirow{2}{*}{Points2Sound} & depth & 0.016 & 0.080 & 0.122 & \textbf{0.082} & 0.885 & 1.736 & 0.072 & 0.901 \\
 & rgb-depth & \textbf{0.015} & \textbf{0.073} & \textbf{0.114} & 0.099  & \textbf{0.762} & \textbf{1.521} & \textbf{0.067} & \textbf{0.794} \\
 \hline
\end{tabular}
\label{table:quantitative_results_m2b}
\end{table*}

In the appendix, we show a quantitative comparison of Points2Sound ($\mathcal{L}_{\mathrm{full}}$) with a recent spectrogram-based Mono2Binaural model from Gao et al.~\cite{gao20192}.  Mono2Binaural was designed to generate a binaural version from mono audio at 16 kHz using 2D visual information as guidance. In this case, we adapt it for audio recordings sampled at 44.1 kHz and 3D visual information as guidance. The original Mono2Binaural extracts visual features using a Resnet18 with dense convolutions while audio features and audio-visual analysis is performed using a U-Net. We use the same sparse Resnet18 from Points2Sound to extract the visual feature from the 3D scene. In addition, in order to resemble its original model, the last Resnet18 3x3x3 sparse convolution is implemented with $K=512$ channels. Then, as in Mono2Binaural, the visual feature vector is replicated to match the spatial feature dimensions of the U-Net bottleneck and concatenated along the channel dimension. During training, we select 0.63 seconds clips of audio and compute the STFT using a Hann window of 23 milliseconds and hop length of 10 milliseconds. Mono2Binaural considers mono inputs represented as $s_{m} = s_{b}^L+s_{b}^R$ and the learning objective is to predict the complex-valued spectrogram of the difference of the two binaural channels. Then, both predicted binaural channels are recovered as follows:
\begin{equation}
\hat{s_{b}}^L = (s_{m}+\hat{s_{b}}^{\mathrm{diff}})/2, \quad \hat{s_{b}}^R = (s_{m}-\hat{s_{b}}^{\mathrm{diff}})/2 
\end{equation}
where $s_{b}^{\mathrm{diff}}(t)= s_{b}^{L}-s_{b}^{R}$. We use Adam optimizer and minimize the mean squared error loss function. During testing, Mono2Binaural uses a sliding window with a hop size of 50 milliseconds to binauralize the 10 seconds audio clips. 

Figure \ref{fig:box_plot_results} shows a visual comparison of the performance of Points2Sound and Mono2Binaural for different sources when rgb-depth point cloud features are used. Table \ref{table:quantitative_results_m2b} shows quantitative results of both learning methods for different types of 3D point cloud attributes and number of sources.

In addition, we provide a second supplementary video with listening examples where three audio-visual scenes from the test set with $N=2$ sources are present. For each listening example, we first show the 3D point cloud scene, and then provide the input mono audio, the Points2Sound and Mono2Binaural predicted binaural audios, and the ground truth binaural audio. The audio-visual scenes are selected to contain sound sources that are not located in the same side of the listener's head. Also, the scenes contain a variety of sound sources which play in the same frequency range in some fragments.

\section*{Acknowledgments}
The authors thank Alexander Mayer for technical support. We also thank the reviewers for comments leading to significant improvements. This project has received funding from the European Union's Horizon 2020 research and innovation programme under the Marie Sk\l{}odowska-Curie grant agreement No 812719.

\section*{Author Contributions}
\textbf{FL} proposed the idea, wrote the code, ran the experiments and wrote the paper. \textbf{VC} and \textbf{AH} supervised the research.

{\small
\bibliographystyle{ieee_fullname}
\bibliography{egbib}
}

\end{document}